\documentclass{aa}
\usepackage{txfonts}
\usepackage{natbib}
\bibpunct{(}{)}{;}{a}{}{,}
\usepackage[dvips]{graphicx}

\newcommand{\pun}[1]{\mbox{\rm\,#1}} % Command used to write physical units

\newcommand{\logg}{\ensuremath{\log g}}

\newcommand{\moh}{\ensuremath{[\mathrm{Fe/H}]}}

\newcommand{\Teff}{\ensuremath{T_{\mathrm{eff}}}}

\newcommand{\beq}{\begin{equation}}
\newcommand{\eeq}{\end{equation}}

\newcommand{\eref}[1]{\mbox{(\ref{#1})}}
\newcommand{\ebref}[1]{\mbox{[\ref{#1}]}}
\newcommand{\plabel}[1]{\label{#1}}

\newcommand{\COROT}{{\sf CoRoT}}
\newcommand{\COBOLD}{{\sf CO$^5$BOLD}}
\newcommand{\ATLAS}{{\sf ATLAS}}
\newcommand{\MARCS}{{\sf MARCS}}
\newcommand{\VIRGO}{{\sf VIRGO}}
\newcommand{\SOHO}{{\sf SOHO}}

\newcommand{\mohi}{S0}
\newcommand{\molo}{S1}
\newcommand{\mosu}{Sun1}

\newcommand{\Hpsurf}{\ensuremath{H_\mathrm{P}^\mathrm{surf}}}

\newcommand{\dirms}{\ensuremath{\delta I_\mathrm{rms}/I}}
\newcommand{\tauc}{\ensuremath{\tau_\mathrm{c}}}
\newcommand{\nobm}{\ensuremath{\mathrm{N}_\mathrm{obm}}}

\newcommand{\imu}{\ensuremath{m}}

\newcommand{\nmu}{\ensuremath{M}}

\newcommand{\Ntot}{\ensuremath{N}}
\newcommand{\Iobs}{\ensuremath{I}}
\newcommand{\Imu}{\ensuremath{I_{\imu}}}

\newcommand{\HImu}{\ensuremath{\FT{I}_{\imu}}}

\newcommand{\HCImu}{\ensuremath{\conj{\FT{I}}_{\imu}}}

\newcommand{\mueffmu}{\ensuremath{\mu_{\imu}}}

\newcommand{\var}[1]{{\ensuremath{\sigma^2_{#1}}}}
\newcommand{\sig}[1]{{\ensuremath{\sigma_{#1}}}}

\newcommand{\lx}{\ensuremath{l_\mathrm{x}}}
\newcommand{\ly}{\ensuremath{l_\mathrm{y}}}
\newcommand{\lz}{\ensuremath{l_\mathrm{z}}}
\newcommand{\lgran}{\ensuremath{l_\mathrm{gran}}}
\newcommand{\siggran}{\ensuremath{\sig{\mathrm{gran}}}}
\newcommand{\siggransq}{\ensuremath{\sigma^2_\mathrm{gran}}}
\newcommand{\fobs}{\ensuremath{f}}
\newcommand{\Rstar}{\ensuremath{R}}
\newcommand{\Rsun}{\ensuremath{\mathrm{R}_{\sun}}}

\newcommand{\msum}[1]{\ensuremath{\sum_{#1=1}^{\nmu}}}

\newcommand{\wmu}{\ensuremath{c_\imu}}

\newcommand{\tmean}[1]{\ensuremath{\left\langle #1\right\rangle}}

\newcommand{\xmean}[1]{\ensuremath{\overline{#1}}}

\newcommand{\Hfobs}{\ensuremath{\FT{\fobs}}}
\newcommand{\HCfobs}{\ensuremath{\conj{\FT{\fobs}}}}
\newcommand{\FT}[1]{\ensuremath{\hat{#1}}}
\newcommand{\conj}[1]{\ensuremath{#1^\ast}}

\newcommand{\draftflag}{false}

\begin{document}

\title{Hydrodynamical simulations of convection-related stellar micro-variability}
\subtitle{II. The enigmatic granulation background of the \COROT\ target
  HD\,49933\thanks{The \COROT\ space mission, launched on December 27th 2006,
    has been developed and is operated by CNES, with the contribution of
    Austria, Belgium, Brazil, ESA, Germany and Spain.}}

% For referee layout:
\titlerunning{The enigmatic granulation background of the \COROT\ target HD\,49933}
%\titlerunning{}
\authorrunning{Ludwig et al.}

\author{ H.-G. Ludwig\inst{1,2} \and
R. Samadi\inst{3} \and 
M. Steffen\inst{4} \and
T. Appourchaux\inst{5} \and
F. Baudin\inst{5} \and
K. Belkacem\inst{6} \and
P. Boumier\inst{3} \and
M.-J. Goupil\inst{3} \and 
E. Michel\inst{3}
}

\institute{
GEPI, Observatoire de Paris, CNRS, Univ. Paris 7,
F-92195 Meudon Cedex, France; \email{Hans.Ludwig@obspm}
\and
CIFIST Marie Curie Excellence Team
\and
LESIA, Observatoire de Paris, CNRS (UMR 8109), Univ. Paris 6, Univ. Paris 7,
F-92195 Meudon Cedex, France
\and
Astrophysikalisches Institut Potsdam, An der Sternwarte 16, D-14482 Potsdam, Germany
\and
Institut d'Astrophysique Spatiale, Univ. Paris 11, CNRS (UMR 8617), F-91405 Orsay, France
\and
Institut d'Astrophysique et de G{\'e}ophysique de l'Universit{\'e} de
Li{\`e}ge, All{\'e}e du 6 Ao{\^u}t 17, B-4000 Li{\`e}ge, Belgium
}

\date{Received ???; accepted ???}

  \abstract
  {Local-box hydrodynamical model atmospheres provide statistical information
    about a star's emergent radiation field which allows one to predict the level
    of its granulation-related micro-variability. Space-based photometry is
    now sufficiently accurate to test model predictions.}
{We aim to model the photometric granulation background of HD\,49933 as
  well as the Sun, and compare the predictions to the measurements obtained by the \COROT\
  and \SOHO\ satellite missions.}
   {We construct hydrodynamical model atmospheres representing HD\,49933 and
     the Sun, and
     use a previously developed scaling technique to obtain the observable disk-integrated
     brightness fluctuations. We further performed exploratory
     magneto-hydrodynamical simulations to gauge the impact of small scale
     magnetic fields on the synthetic light-curves.}
   {We find that the granulation-related brightness fluctuations depend on
     metallicity.  We obtain a satisfactory correspondence between prediction
     and observation for the Sun, validating our approach. For HD\,49933, we
     arrive at a significant over-estimation by a factor of two to three in
     total power.  Locally generated magnetic fields are unlikely to be
     responsible, otherwise existing fields would need to be rather strong to
     sufficiently suppress the granulation signal.  Presently suggested
     updates on the fundamental stellar parameters do not improve the
     correspondence; however, an ad-hoc increase of the HD\,49933 surface
     gravity by about 0.2\pun{dex} would eliminate most of the discrepancy.}
     {We diagnose a puzzling discrepancy between the predicted and observed granulation background
      in HD\,49933, with only rather ad-hoc ideas for remedies at hand.}

\maketitle

\section{Introduction}

The bright ($m_\mathrm{V}=5.77$) F-type dwarf HD\,49933 (HR\,2530, ADS~5505A)
was the first prime seismic target of the \COROT\ space photometry mission
\citep{Auvergne+al09}. A clear signature of solar-like oscillations was
detected in the acquired light-curve \citep{Appourchaux+al08,Michel+al08}. This
confirms expectations about the presence of solar-like oscillations in
HD\,49933 which were anticipated from results of ground-based observations of
radial velocity \citep{Mosser+al05}.

It is now generally accepted that solar-like oscillations are excited by
stochastic convective motions taking place in the stellar surface layers on
time-scales of the same order as the periods of the excited modes.  The
convective flows are thermally driven, and imprint a time-varying brightness
pattern -- granulation -- on the stellar surface.  While the oscillations
manifest themselves as discrete peaks in a temporal power spectrum of a star's
light-curve, the stochastic evolution of the granulation pattern leads to a
continuous signal, the so-called granulation background, since it usually
constitutes the background on which oscillatory peaks are superimposed in a
spectrum.  Oscillations, as well as granulation, induce very small brightness
fluctuations, challenging the sensitivity level of photometry achievable from
the ground, and motivating photometry from space such as that performed by
\COROT.

The modelling of stellar granulation -- mainly driven by the need for improved
stellar atmosphere models of late-type stars -- made substantial progress over
the last two decades. It has now reached a level that should allow a direct
comparison between the predicted and observed granulation properties, in
particular the granulation background
\citep{Trampedach+al98,Svensson+Ludwig05,Ludwig06,Guenther+al08}. Hitherto,
the comparison was somewhat hampered by the limited photometric accuracy, duty
cycle, and duration of existing stellar measurements.  The outstanding
accuracy achieved on HD\,49933 by \COROT\ offers a new opportunity for a
stringent comparison between predicted and observed granulation background,
and is attempted here. We will see, however, that some problems are encountered.

The paper is organized as follows: we briefly describe the photometric
data (Sect.~\ref{s:photo}), and the hydrodynamical granulation models
(Sect.~\ref{s:models}), comment on how we derive the synthetic background
spectra (Sect.~\ref{s:synspec}), present comparisons for the Sun and
HD\,49933 (Sect.~\ref{s:results}), and finish by discussing effects which may
play a role in the encountered problems (Sects.~\ref{s:discussion}
and~\ref{s:conclusions}).

\section{\COROT\ photometric data\plabel{s:photo} and power spectra}

We used reduced photometric data of the N2-level \citep{Samadi+al07} covering
137\pun{d} for HD\,49933 from long run LRa01, and 157\pun{d} for the F-dwarfs
HD\,181420 and HD\,181906 from long run LRc01. We applied the same bolometric
correction \citep{Michel+al08b} to all targets, amounting to a factor of 1.11
in amplitude of the brightness fluctuations. We calculated power spectra
by applying a standard Fast Fourier Transform from the bolometric light-curve
without further processing. In particular, we did not apply any correction to
the long-term trend apparent in the light-curve of HD\,49933. The reason was
that applying the measured loss of instrumental sensitivity due to aging
again produced a trend but in the opposite sense. This in turn resulted in a
power spectrum almost indistinguishable from the power spectrum of the
uncorrected data. On the other hand, simply de-trending by removing any linear
long-term variation had an impact on the power -- but at frequencies $\nu <
0.3\pun{mHz}$ only. These frequencies are not relevant to the conclusions
drawn in this paper. Hence, we did not further investigate this issue, but
remark that there is an instrument-related uncertainty in the measured power
and its slope in the range $\nu < 0.3\pun{mHz}$.

When we refer to ``power'' in this paper we always mean power
spectral density. We follow the normalization that the power
integrated over the interval between zero (excluded) and the Nyquist frequency
equals the variance \var{} of the signal in the time domain.

\section{\COBOLD\ radiation-hydrodynamics simulations\plabel{s:models}}

We used the radiation-hydrodynamics code \COBOLD\ \citep[for further
information about the code and applications see ][]{Freytag+al02,Wedemeyer+al04}
to construct three 3D Cartesian ``local-box'' model atmospheres. Two models were
intended to represent HD\,49933, and one -- for reference -- the Sun.
Table~\ref{t:models} summarizes the model properties. The atmospheric
parameters for HD\,49933 were chosen close to values recommended in the
literature, $\Teff=6780\pm 130\pun{K}$ \citep{Bruntt+al08}, and
$\logg\approx 4.25$ \citep{Appourchaux+al08}. The metallicity of the star
is $\moh\approx -0.37\pun{dex}$ \citep{Solano+al05,Gillon+Magain06}. The
opacities used in the models for HD\,49933 are based on detailed
low-temperature opacities from the \MARCS\ stellar atmosphere package
\citep{Gustafsson+al08}. We only had these data available for particular
metallicities, and not for arbitrary element mixtures. Due to this
restriction, we decided to compute two 3D models with metallicities which
bracket the star's observed metallicity. As will be demonstrated below, the
chemical composition has an important impact on the model properties. The
solar model~\mosu\ is a \COBOLD\ model of an older generation employing
opacities from the \ATLAS\ model atmosphere package \citep{Kurucz05}. While
this leads to some systematic differences between the models, we believe that
they so not influence the comparisons performed in this paper.
Our two hydrodynamical model
atmospheres for HD\,49933 were also employed to predict excitation
rates of oscillatory modes \citep{Samadi+al08,Samadi+al09}.

Unfortunately, the aforementioned stellar parameters are not completely
reliable. \citet{Bruntt+al04} list effective temperatures and surface
gravities from various determinations which differ by almost 500\pun{K} and
0.5\pun{dex}, and the debate about the exact parameters is still ongoing
\citep{Bruntt09,Kallinger+al08}. It may turn out that HD\,49933 is
200--300\pun{K} cooler than assumed in our models. More important for the
level of the granulation background, however, is the the value of the surface
gravity.  We will come back to this point later but want to emphasize that
HD\,49933 has an accurately known parallax of $33.7\pm0.4\pun{mas}$ from {\sf
  HIPPARCOS} \citep{Leeuwen07} which helps to constrain its surface gravity
and stellar radius. In this paper, we use a radius of 1.35\pun{\Rsun} for
HD\,49933, within the error box given by \citet{Appourchaux+al08} of $1.36\pm
0.06\pun{\Rsun}$.

\begin{table*}

\caption{\COBOLD\ radiation-hydrodynamics model atmospheres: ``Model'' is the
    model's name used in this paper, \Teff\ the effective temperature,
    \logg\ the gravitational acceleration, \moh\ the metallicity, \Rstar\ an assumed stellar radius
    (not intrinsic to the simulation proper), $\lx=\ly$ the linear horizontal size of
    the square-shaped computational box, \lz\ its vertical extent, 
    $T$ the duration of the simulated time series,
    \tauc\ the sound crossing time over \Hpsurf,
    \Hpsurf\ the pressure scale height at Rosseland optical depth unity, \dirms\
    the relative spatial white light intensity contrast at disk-center,
    \siggran\ the temporal relative disk-integrated
    granulation-related brightness fluctuations in white light 
    (in parenthesis an estimate of the uncertainty),
    \nobm\ the number of
    equivalent frequency points considered in the solution of the radiative
    transfer equation, ``Modelcode'' an internal identifier of the model
    sequence.
    \label{t:models}}
\begin{center}
\begin{tabular}[t]{lllllllllllllll}
\hline\noalign{\smallskip}
Model &  $T_\mathrm{eff}$ & \logg  & \moh & \Rstar & \lx,\ly & \lz & $T$ 
& $\tau_\mathrm{c}$ & $H_{\mathrm{p}}^{\mathrm{surf}}$ & \dirms & \siggran & \nobm & Modelcode\\
& [K] & [$\mathrm{cm/s^2}$] & & [\Rsun] & [Mm] & [Mm] & [ks] & [s] & [Mm] & &[ppm]   &  & \\
\noalign{\smallskip}
\hline\noalign{\smallskip} 
\mohi     & 6\,720  & 4.25 & \phantom{-}0.0 & 1.35 &           16.4 & 24.2           &           135.6 & 29.5 & 0.241 & 0.196 & 84 (17)& 5 & d3t68g43mm00n01\\
\molo     & 6\,730  & 4.25 &           -1.0 & 1.35 &           16.1 & 24.2           & \phantom{1}73.8 & 29.1 & 0.238 & 0.229 & 64 (13)& 6 & d3t68g43mm10n01\\
\mosu     & 5\,780  & 4.44 & \phantom{-}0.0 & 1.00 & \phantom{1}5.6 & \phantom{2}2.5 & \phantom{1}52.2 & 17.8 & 0.141 & 0.172 & 40 (8) & 5 & d3gt57g44n53\\
\noalign{\smallskip}
\hline
\end{tabular}
\end{center}
\end{table*}

Since the brightness fluctuations which are central to the present
  investigation are intimately linked to the heating and cooling provided by
  the radiation field to the gas, we give here some details about the treatment of
  the radiative transfer in the hydrodynamical models. All models employ
  $140\times 140 \times 150$ grid points for the spatial mesh in the x-, y-, and
  (vertical) z-direction, respectively. For all models, the mesh spacing is
  equidistant in the horizontal directions. In vertical direction, \mohi\ and
  \molo\ employ a non-equidistant spacing, \mosu\ again an equidistant mesh.  In the
  important layers around $\tau=1$ the spacing in vertical direction amounts
  to 46.9\,km (0.19\,\Hpsurf, definition see Table~\ref{t:models}), 48.2\,km
  (0.20\,\Hpsurf), and 15.1\,km (0.11\,\Hpsurf) for models \mohi, \molo, and
  \mosu, respectively.  Note that the models \mohi\ and \molo\ use the same
  vertical mesh. The difference in resolution at $\tau=1$ is the result of
  how the model adjusts on the computational mesh. The radiative
  transfer is solved along long characteristics employing Feautrier's method
  \citep[e.g.,][]{Mihalas78} assuming strict LTE.  Corresponding ``rays'' are
  started at each point of the spatial mesh in the top-most layer. All models
  use bundles of rays in one vertical and two inclined directions, as well as
  four azimuthal directions \citep[$M=3$ case described in][]{Ludwig06}
  coinciding with the coordinate axes. In total this amounts to 176\,400 rays
  along which the radiative transfer equation is solved at each time step.
  The number of equivalent wavelength points~\nobm\ to represent the
  wavelength dependence of the radiation field is five for models \mohi, and
  \mosu, six for \molo, as given in Table~\ref{t:models}. The necessary
  interpolation of the radiative heating or cooling between the system of rays
  and spatial mesh (for hydrodynamics) is performed in an energy-conserving
  fashion. To optimize performance, the treatment of the
  radiative transfer is switched to the diffusion approximation in the deep,
  optically thick layers.

Finally, we want to point out two aspects which are of particular importance
  for the interpretation of the granular background. Qualitatively, the lower
  metallicity in model \molo\ in comparison to \mohi\ leads to a lower overall
  opacity (primarily due to the lower H$^-$ opacity at lower electron
  pressure). This in turn implies an increase of the mass density at given
  optical depth. At optical depth unity, we find an increase of the mass
  density by $\approx 40\pun{\%}$ in model \molo\ relative to \mohi. At fixed
  flow geometry, the higher density would require a smaller temperature
  contrast or smaller velocity differences between up- and down-flowing
  material to transport the prescribed stellar energy flux. What actually
  happens in \molo\ is an increase of the temperature contrast reflected by an
  increase of the intensity contrast (see Table~\ref{t:models}). This is
  likely a consequence of the dependence of the opacity on temperature. On the
  other hand, as expected, the typical velocity amplitudes in model \molo\ are
  smaller by about $\approx 10\pun{\%}$ at the maximum of the vertical
  velocity. Finally, the typical granular scales become shorter in \molo.  The
  combination of these factors determines the amplitude and characteristic
  frequency of the granulation background.

\subsection{Synthetic power spectra of the observable flux\label{s:synspec}}

The hydrodynamical models provide time series of the radiation intensity at
different limb-angles including the temporal convection-related fluctuations.
\citet{Ludwig06} showed that, together with an assumption about the stellar
radius, this information is sufficient to predict the power
spectrum of the disk-integrated, observable, relative
brightness variations. He obtained for the frequency component of the power
spectrum the relation
\beq
\frac{\tmean{\Hfobs\HCfobs}}{\tmean{\fobs}^2}=
\Ntot^{-1}\frac{\msum{\imu}\wmu\,\mueffmu^2\tmean{\HImu\HCImu}}%
{\left(\msum{\imu}\wmu\,\mueffmu\tmean{\Imu}\right)^2}
\plabel{e:j1power2}
\eeq
where 
\beq
N = \frac{2\pi \Rstar^2}{A}.
\eeq
Angular brackets denote expectation values, \fobs\ the observable
disk-integrated flux, and \Hfobs\ its (complex) Fourier transform. The asterisk
indicates the conjugate complex, \wmu\ a discrete integration weight,
\mueffmu\ the cosine of the limb-angle, and \Imu\ the spatial average of the
intensity in direction cosine~\mueffmu. According Eq.~\eref{e:j1power2} the
observable power scales inversely proportional to the number~\Ntot\ of
simulation boxes of surface area~$A$ tiling the visible stellar hemisphere of
radius~\Rstar. For a detailed derivation and discussion of
Eq.~\eref{e:j1power2} see \citet{Ludwig06}.

Due to the finite duration of a simulation, the obtained power spectra of the
relative brightness fluctuations exhibit appreciable noise. To improve the
effective signal-to-noise level and facilitate the comparison to observations
we fit a simple analytical model to the simulated spectra. The
analytical model reflects our prejudice that the frequency dependence of the
granulation-related background signal should show rather little structure.
Moreover, it allows us to eliminate the acoustic eigenmodes of the
computational domain which are excited by the convection-related fluctuations
but are not directly comparable to the observed modes. We applied a model for
the spectral power density~$P$ which is a generalized Harvey model
\citep{Harvey85} for the background, plus a sum of Lorentzians for the
box-modes according to
\beq
P(\nu) = \frac{b}{1+\left(\frac{\nu}{\nu_1}\right)^{\alpha_1}+\left(\frac{\nu}{\nu_2}\right)^{\alpha_2}}
+ \sum_{k=1}^K \frac{a_k}{1+\left(\frac{\nu-u_k}{w_k}\right)^2}. 
\plabel{e:fitmodel}
\eeq
$\nu$ is the cyclic (temporal) frequency. All other variables are fitting
parameters of the model: $b$ is the asymptotic amplitude of the background
towards low frequencies, $\nu_1$ and $\nu_2$ characteristic frequencies of the
background, $\alpha_1$ and $\alpha_2$ power law exponents, $a_k$ are the mode
amplitudes of $K$ considered modes, $u_k$ their frequency positions, $w_k$
their line widths. We performed a maximum likelihood estimation of the
parameters; we used the commonly adopted model that each frequency component
of a power spectrum is statistically independent and follows a probability
distribution~$p$ which is a $\chi^2$-distribution with two degrees of freedom.
This is an exponential of the form
\beq
p(x) = \frac{1}{\tmean{x}}\exp\left(-\frac{x}{\tmean{x}}\right).
\label{e:expdist}
\eeq
\tmean{x} is the expectation value of the random variable~$x$. The
model~\eref{e:fitmodel} allows us to extrapolate the fitted spectrum to
frequencies higher than the Nyquist frequency~$\nu_\mathrm{Nyquist}$. To
mitigate effects of aliasing present at the highest frequencies of the
simulated spectra, we did not fit $P(\nu)$ itself but
$P(\nu)+P(2\nu_\mathrm{Nyquist}-\nu)$. This includes
the ``mirroring'' of high frequency power into the frequency domain of
interest by aliasing. As we will see, $P(\nu)$ decreases strongly towards high
frequencies so that aliasing effects are only noticeable close to the Nyquist
frequency.

To maximize the likelihood function we used the {\sf POWELL} function
implemented in {\sf IDL}. The maximization problem turned out not to be well
conditioned, and some manual intervention was always necessary to obtain a
stable and acceptable fit. This points to the need for a more sophisticated
fitting procedure with suitable regularization. However, for the present
purposes we considered the fitting rather a kind of constrained smoothing
so that the shortcomings from the viewpoint of statistical stringency were not
considered vital. We verified that our fitting resulted in a representation
close to what was obtained by direct smoothing of the simulated spectra.

\section {Results\plabel{s:results}}

\subsection{Fitting the simulated raw spectra}

Figure~\ref{f:fitting} illustrates the result of fitting the analytical model
of Eq.~\eref{e:fitmodel} to the simulated raw spectra calculated from the time
series with the help of Eq.~\eref{e:j1power2}. Table~\ref{t:fitpars} lists the
fitted parameters. As already mentioned, the simulated spectra show an
appreciable level of noise. In the figure, the simulated spectra were smoothed
to allow a clearer comparison. By comparing solar models with different
numerical set-ups, we estimated that the absolute uncertainty in the simulated
spectra amounts to about $\pm 20\pun{\%}$ in total (i.e. frequency integrated)
amplitude; we expect the relative accuracy among models of similar numerical
set-ups to be somewhat better. Note that we left out the contribution of the
box modes in the fitting functions in Fig.~\ref{f:fitting}.  The simulated
spectra show clear differences in the total power between the Sun and
HD\,49933 -- primarily reflecting the change of the granular cell size
relative to the stellar radius among the stars. The metal-depleted model for
HD\,49933~\molo\ shows only 58\% of the model~\mohi\ in total power.  This is
the result of a reduction of the granular scale at lower metallicity,
over-compensating of the increase in the contrast (cf. Eq.~\ebref{e:dfof}).
Moreover, the power in model~\molo\ is shifted towards lower frequencies.
This we trace back to longer granular life-times in \molo, presumably related
to the lower typical velocities in \molo.  All models predict a rather steep
drop of the granular background signal towards high frequencies, similar to the
one observed in the Sun.

\begin{table}
\begin{center}
  \caption[]{%
    Fitted parameters of Eq.~\eref{e:fitmodel} for the synthetic power spectra
    of the three hydrodynamical models.  Background- and line-amplitudes are
    given in ppm$^2$/$\mu$Hz, frequencies in mHz. Missing values for a line
    indicate that it was left out since unnecessary to obtain a satisfactory
    fit.  \plabel{t:fitpars}}
\begin{tabular}{llll}
\hline\noalign{\smallskip}
Parameter  & \multicolumn{3}{c}{Hydrodynamical models}\\
\mbox{}    & \mohi & \molo & \mosu \\
\noalign{\smallskip}
\hline\noalign{\smallskip}
$b$        & 6.056     & 2.881   & 1.046 \\
$\nu_1$    & 0.9593    & 1.259   & 1.295 \\
$\alpha_1$ & 2.486     & 3.434   & 2.953 \\
$\nu_2$    & 2.383     & 2.821   & 3.667 \\
$\alpha_2$ & 8.061     & 9.278   & 8.035 \\ 
$a_1$      & 0.4130    & 0.5389  & 2.149 \\
$u_1$      & 1.266     & 1.674   & 3.252 \\
$w_1$      & 0.09507   & 0.08464 & 0.02861 \\
$a_2$      & 4.498     & 1.732   & \mbox{}\\
$u_2$      & 1.676     & 2.027   & \mbox{}\\
$w_2$      & 0.01155   & 0.02994 & \mbox{}\\
$a_3$      & 2.188     & 2.142   & \mbox{}\\
$u_3$      & 2.107     & 2.618   & \mbox{}\\
$w_3$      & 0.01217   & 0.004683& \mbox{}\\
\noalign{\smallskip}
\hline
\end{tabular}%
\end{center}
\end{table}

\begin{figure}
\begin{center}
\resizebox{\hsize}{!}{\includegraphics[draft=\draftflag,angle=90]{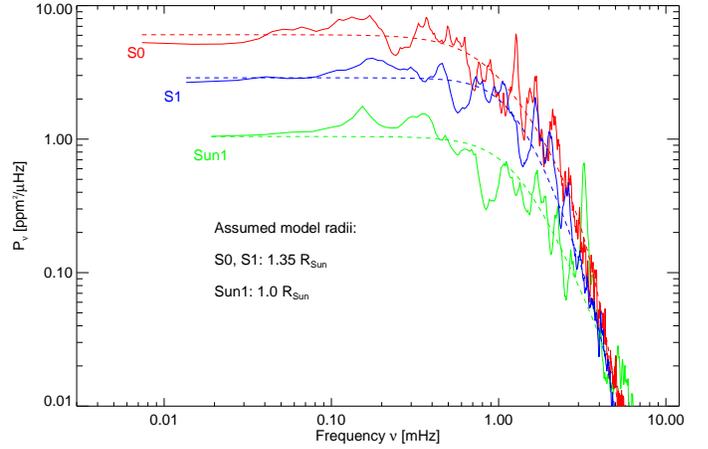}}
\caption{Inter-comparison of the raw (solid lines) and fitted (dashed lines)
  spectra (power spectral density) of the three hydrodynamical simulations. The contribution of
  box-modes to the fitted signal was left out in the graphs.\plabel{f:fitting}}
\end{center}
\end{figure}

\subsection{Comparison with \VIRGO\ observations of the Sun}

Figure~\ref{f:virgocomp} constitutes an update of results presented by
\citet{Svensson+Ludwig05}, and shows a comparison (not fit) between the
disk-integrated, photometric fluctuations derived from the solar model~\mosu,
and observational data from the \VIRGO\ instrument on board the \SOHO\
satellite.  Our focus is the high frequency region of the solar signal in
which granular contributions dominate. The \VIRGO\ power spectrum was
calculated from (level 2) time series data provided by the \VIRGO\ team.
One year of data (1996.5-1997.5) was chosen close to solar minimum
activity to minimize the possible contribution of activity related
variability. The plotted power spectrum is based on data of the green channel
of the \VIRGO\ three-channel sun-photometer (SPM, see \cite{Froehlich+al97}).
Following the procedure of \citet{Svensson+Ludwig05}, it has been converted to
white light fluctuations by matching (by shifting in power) a corresponding
power spectrum based on \VIRGO\ PMO6V-A absolute radiometer data in the
frequency range 0.3 to 2.0\pun{mHz}. Instead of using the PMO6V-A power
spectrum directly, this rather involved procedure was necessary since PMO6V-A
and SPM power spectra deviate substantially in the high frequency region. The
authors found little information about this mismatch in the literature.
However, it appears to be agreed that the SPM spectrum reflects the actual
solar behavior \citep{Froehlich+al97,Andersen+al98}, in particular showing
the steep (roughly as $\nu^{-4}$) decline at the highest frequencies.

\begin{figure}
\begin{center}
\resizebox{\hsize}{!}{\includegraphics[draft=\draftflag,angle=90]{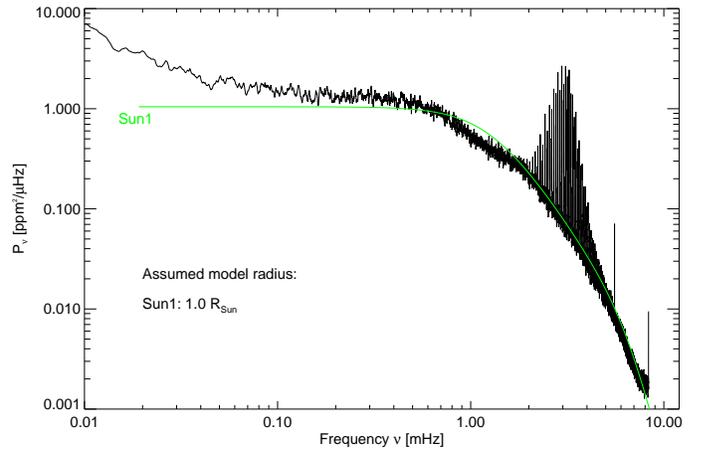}}
\caption{Power spectra of disk-integrated photometric
  fluctuations for the Sun: the predicted background signal of model~\mosu\
  (green/grey solid line) and observational data from \SOHO/\VIRGO\ (black
  solid line). Note the steep decline in power towards high
  frequencies.\plabel{f:virgocomp}}
\end{center}
\end{figure}

We find a satisfactory agreement of the continuous background signal between
model predictions and observations between 0.1 and 8.0\pun{mHz}. In
particular, the background in the p-mode frequency range is matched quite
well. The observed background power shows a ``kink'' at frequencies slightly
above 1\pun{mHz} relative to the model prediction, and the ``plateau'' power
around 0.2\pun{mHz} is not exactly represented.  \citet{Michel+al08b}
interpreted the ``kink'' as the sign of the transition between a meso-granular
and granular contribution to the power spectrum. This interpretation is
obviously not supported by our granulation model, which already provides too
much power around 1\pun{mHz}, and has a geometrical dimension which would not
allow the development of meso-granular convection cells of 5--10\pun{Mm} size
\citep{November+al81}. Moreover, Michel and co-workers attribute a similar
total fluctuation amplitude to meso-granulation and granulation.  According to
Eq.~\eref{e:dfof}, this would correspond to a spatial brightness contrast which
would be a sizable fraction of the granulation contrast and should be easily
recognizable in images of the solar surface -- which is not the case.  

Towards low frequencies ($\nu<0.1\pun{mHz}$), the increase of the observed
signal is due to the solar magnetic activity, and cannot be represented by the
pure (non-MHD) hydrodynamical model applied here. We also considered the power
spectrum at solar maximum (year 2000.5-2001.5, not shown). Towards solar
maximum, the activity-related power increases but leaves the
granulation-related signal intact; the power spectrum is essentially identical
at frequencies $\nu > 0.2\pun{mHz}$.  It is difficult to provide an objective
criterion to determine to which lowest temporal frequency our model can make reliable
predictions. A limit is set by the maximum size that features can develop in
the simulation. However, it is not obvious how this relates to their
life-time, hence the frequencies on which they will have an impact.

A conspicuous feature exhibited by the observations as well as the model is
the ``step-like'' shape of the granular background signal. As it will be shown
later, this feature is not obvious in the observations of HD\,49933, while it
is predicted by the simulations.

\subsection{Comparison with \COROT\ observations of HD\,49933\plabel{s:pcomp}}

Figure~\ref{f:phd49933} shows a comparison between the predicted power spectra
(power spectral density) of models \mohi\ and \molo. Ideally, the observed
spectrum should fall somewhere between models \mohi\ and \molo.
This is not at all the case; the predictions lie noticeably higher with a power
almost a factor of two greater than observed around 0.9\pun{mHz}. Moreover, the
overall shape of the observed spectrum is not well represented by the models.
In particular, the conspicuous drop of the background level is not obvious in
the observations. 

In Fig.~\ref{f:phd49933} we added two spectra of F-dwarfs which were also
observed by \COROT. The stars have parameters not too different from HD\,49933
(HD\,181420: \Teff=6650\pun{K}, \logg=4.17, [M/H]=-0.04; HD\,181906:
6380/4.15/-0.14; see \citet{Michel+al08}). The overall levels of the
brightness fluctuations differ, however, the spectra show a remarkable
similarity in shape.  We take this as an indication that HD\,49933 is a
typical representative of its class, and not an exceptional case among
F-dwarfs. We note here that the similarity of the spectral shape at
$\nu>0.3\pun{mHz}$ might be connected to the rather similar rotational
periods of the stars of 3.4\pun{d} for HD\,49933 \citep{Appourchaux+al08},
2.8\pun{d} for HD\,181906 \citep{Garcia+al09}, and 3.5$\ldots$4.5\pun{d} for
HD\,181420 \citep{Barban+al09}.

\begin{figure}
\begin{center}
\resizebox{\hsize}{!}{\includegraphics[draft=\draftflag,angle=90]%
{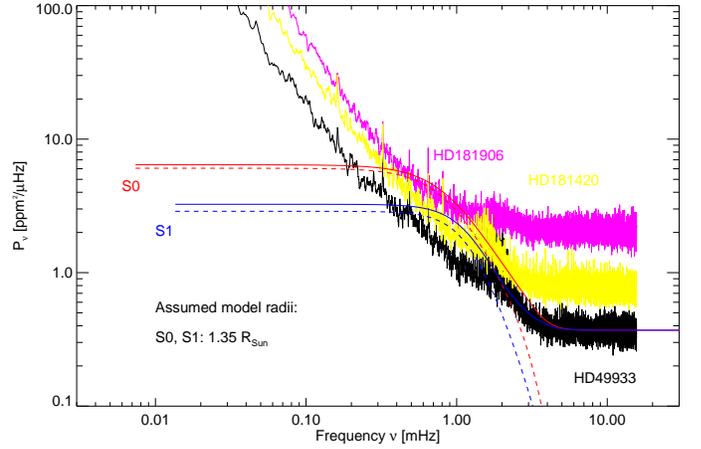}}
\caption[]{%
 Power spectra for HD\,49933: dashed lines depict the predictions from models
  \mohi\ and \molo. The solid lines are again the predicted spectra but
  increased by a constant corresponding to the observed photometric white
  noise level. The black line depicts the spectrum of HD\,49933. The other two
  solid lines depict spectra of the similar F-dwarfs HD\,181420 and
  HD\,181906.\plabel{f:phd49933} } % end of caption
\end{center}
\end{figure}

\citet{Michel+al08} assumed that the continuum in the p-mode region indeed
corresponds to the granulation background, fitted it with a standard Harvey
model, and arrived at granulation-related brightness fluctuations of
$\siggran\approx 40\pun{ppm}$ with appreciable uncertainty.  This would put
the fluctuations in HD\,49933 on the same level as observed in the Sun, and is
significantly lower than our predictions. However, accepting for the moment a
much lower level of fluctuations, we experimented in an ad-hoc fashion by
adding (besides photometric noise) an activity-like signal (assuming a
power-law) to our simulated granulation spectra, and simultaneously scaled
their power and characteristic frequency to enforce correspondence with the
observations.  Figure~\ref{f:pscaled} illustrates the outcome after scaling
the power derived from the models by a factor of~0.33 and the frequency by a
factor of~1.6. This resulted in a plausible -- albeit not entirely
satisfactory -- correspondence. Later we shall try to interpret the scaling in
terms of a change of fundamental stellar parameters (see
Sect.~\ref{s:fupars}).

\begin{figure}
\begin{center}
\resizebox{\hsize}{!}{\includegraphics[draft=\draftflag,angle=90]%
{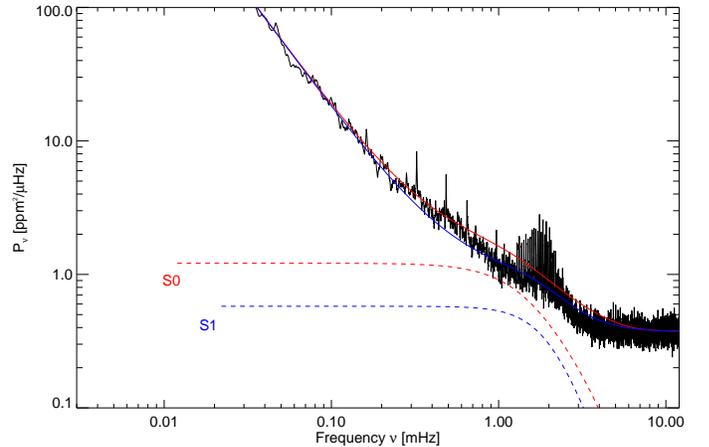}}
\caption[]{%
  Like Fig.~\ref{f:phd49933}, but with scaled models \mohi\ and \molo\
  (for details see text) and adding an ad-hoc signal for magnetic activity and
  photometric noise (solid lines). The dashed lines depict the models after
  scaling without added ad-hoc components.
  \plabel{f:pscaled}} % end of caption
\end{center}
\end{figure}

\section{Discussion\plabel{s:discussion}}

Can we find a reason for the mismatch between model and observation for
HD\,49933? In the following we discuss the impact of uncertainties in the
fundamental parameters and the influence of magnetic activity on the mismatch.

\subsection{Uncertainties in the fundamental parameters\plabel{s:fupars}}

As alluded to before, the fundamental parameters of HD\,49933 (in particular
\Teff, \logg, and \Rstar) might be noticeably different from the values we
assumed in our modeling. \citet{Ludwig06} give an approximate relation
between the spatial contrast of the granulation and the photocentric
variability, which can be equivalently stated in terms of the relative
temporal (RMS) brightness fluctuations~\siggran\ as
\beq
\siggran \propto
\frac{\lgran}{\Rstar}\,\frac{\delta\Iobs_\mathrm{rms}}{\xmean{\Iobs}}.
\plabel{e:dfof}
\eeq
\lgran\ is the typical linear size of a granular cell,
\Rstar\ the stellar radius, \Iobs\ the (spatially resolved) emergent
intensity, $\delta\Iobs_\mathrm{rms}$ its spatial standard deviation, and
\xmean{\Iobs} its spatial average. We wrote a proportionality only since we
want to investigate the differential functional dependence on the fundamental
parameters here. \citet{Freytag+al97} argue on the basis of numerical
simulations that \lgran\ is proportional to the pressure scale height at the
stellar surface~\Hpsurf
\beq
\lgran \propto \Hpsurf \propto \frac{\Teff}{g}.
\eeq
When writing the last proportionality, we assumed a constant mean
molecular weight. For the total (frequency integrated) power~\siggransq\ of the
brightness fluctuations, it follows that using the fundamental relationship between
luminosity~$L$, radius, and effective temperature
\beq
\siggransq
\propto \frac{\Teff^2}{ g^2 \Rstar^2} \,\left(\frac{\delta\Iobs_\mathrm{rms}}{\xmean{\Iobs}}\right)^2
\propto \frac{\Teff^6}{ g^2 L} \,\left(\frac{\delta\Iobs_\mathrm{rms}}{\xmean{\Iobs}}\right)^2.
\plabel{e:deps}
\eeq
Introducing the luminosity in the second proportionality in Eq.~\eref{e:deps}
was motivated by the notion that the accurately known parallax and visual
magnitude of HD\,49933 essentially fixes its luminosity.
Relation~\eref{e:deps} is not exact since the scaling of the granular size
with stellar parameters is not exactly accounted for; for instance, the
formula does not accurately represent the scaling of~\siggran\ among the
models as given in Table~\ref{t:models}. However, we think it is good enough
to provide an estimate of the differential effects in vicinity of the stellar
parameters we used in our modeling.  As evident from Table~\ref{t:models}, the
granulation contrast does not sensitively depend on atmospheric parameters
among F- and G-dwarfs, making it largely invariant over the discussed interval
of temperatures, gravities, and chemical compositions for HD\,49933. Keeping
the luminosity fixed gives estimates of the change of \siggransq\ by
+0.58\pun{dex} with the recently suggested parameters of
\citet{Kallinger+al08} (\Teff=6450\pun{K}, \logg=3.9), and +0.048\pun{dex}
with the ones of \citet{Bruntt09} (``evolutionary solution''
\Teff=6560\pun{K}, \logg=4.19). Despite the high exponent of \Teff\ in
Eq.~\eref{e:deps}, the changes are primarily driven by the changes in \logg. We
are left with the situation that the suggested parameters would increase the
discrepancy between models and observations when applied in the modeling.
For the set of parameters suggested by Kallinger and co-workers the
deterioration would be rather drastic.

The ad-hoc scaling in power and frequency described in Sec.~\ref{s:pcomp} to
enforce the correspondence to observations was motivated by Eq.~\ref{e:deps},
and by the notion that the granular life-time scales with the atmospheric
acoustic cut-off frequency \citep{Kjeldsen+Bedding95,Svensson+Ludwig05} as
$\nu\propto g/\sqrt{\Teff}$. This allows us to associate the scaling factors in
power and characteristic frequency with changes of -200\pun{K} in \Teff\ and
+0.2\pun{dex} in \logg\ of HD\,49933 (assuming constant luminosity). Again,
while helping to reduce the present discrepancies, in particular the increase
of the gravity is at odds with the latest determinations of the star's fundamental
parameters.

\subsection{Influence of magnetic activity}

By analogy to the rise of power towards low frequencies due to magnetic
activity in the Sun (Fig.~\ref{f:virgocomp}), one is tempted to attribute the
rise of power seen in HD\,49933 to magnetic activity as well. This is quite
plausible, since the light-curve of HD\,49933 exhibits clear signatures of
spottedness \citep{Mosser09}. Moreover, HD\,49933 is a rather rapid rotator,
favoring dynamo action. However, polarimetric observations (with {\sf NARVAL}
and {\sf ESPADONS}) indicate that there is no magnetic field present above
$\approx 1\pun{G}$ which is structured on spatial scales of the order of the
stellar radius \citep{Catala+al09}. This raises the question of what kind of
mechanism could produce the apparently ubiquitous field on smaller scales?

There is mounting theoretical and observational evidence
\citep{Cattaneo99,Trujillo+al04,Voegler+Schuessler07} that dynamo action is
possible in convective surface flows generating magnetic fields on very small
scales. To investigate whether such magnetic fields can explain the puzzling
shape of the spectrum around 1\pun{mHz} in HD\,49933, we performed exploratory
MHD simulations with \COBOLD\ \citep{Steffen+al09}. They were restricted to
two spatial dimensions, and performed for solar atmospheric conditions. We
calculated several runs with different, prescribed magnetic flux levels
corresponding to field-free to weak plage regions.  Figure~\ref{f:bspectra}
shows temporal power spectra of the light curves of the simulation runs (not
scaled to full disk). Despite that the simulations were tailored to solar
conditions, we think that the very similar topology of granular flows among
solar-type stars makes them qualitatively also applicable to HD\,49933. The
power spectra of the light-curves show a trend towards lower power at higher
frequencies for increasing total magnetic flux. Obviously, small scale
magnetic fields alter the convective dynamics on small spatial scales but do
not lead to the formation of structures on large scales which may leave an
imprint at lower frequencies in the brightness fluctuations. The results are
preliminary, and further work is necessary. However, for the time being we
consider it unlikely that local dynamo action in the granular flow -- if at
all present --- can explain the shape of the power spectrum of HD\,49933 in
the p-mode frequency region.

\begin{figure}
\begin{center}
\resizebox{\hsize}{!}{\includegraphics[draft=\draftflag,angle=90]%
{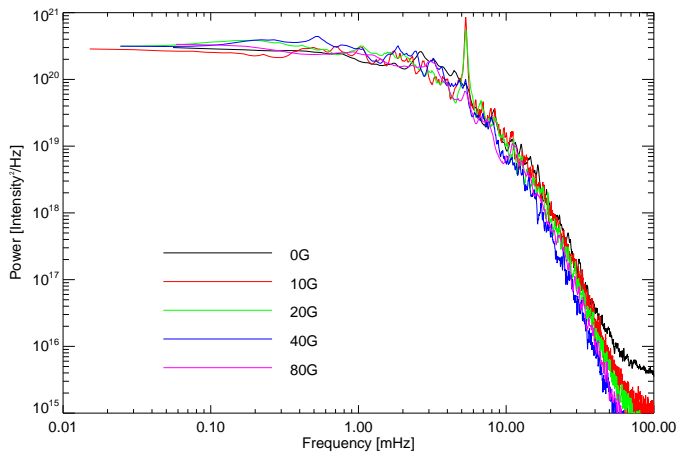}}
\caption[]{% 
Temporal power spectra of the horizontally averaged emergent intensity
in vertical direction of different 2D MHD runs assuming different levels of
initial field strength (solid lines of different colors/shades of grey). The
simulations were performed assuming solar atmospheric parameters.
\plabel{f:bspectra} 
} % end of caption
\end{center}
\end{figure}

Equation~\eref{e:dfof} shows that a reduction of the granular scale at fixed
stellar radius could reduce the observed granulation-related brightness
fluctuations. It is known that magnetic fields of sufficient strength have this
effect \citep[for a recent example see][and references
therein]{Jacoutot+al08}. However, to obtain a sizable reduction one would need
a mean field strength of several 100\pun{G}. A local dynamo appears unlikely
to sustain such magnetic flux levels \citep{Voegler+Schuessler07}, which leaves
us with the need for an efficient generation of magnetic fields on larger
scales.

\section{Conclusions\plabel{s:conclusions}}

3D hydrodynamical model atmospheres can be applied to predict
granulation-related disk-integrated brightness fluctuations. We demonstrated
that the theoretical approach successfully reproduces the
convection-related background signal in the frequency region of the observed
solar p-modes.  Unexpectedly, we found a mismatch between
predicted and observed background when applying the same approach to HD\,49933,
which we argued is a normal representative among the F-dwarfs observed by
\COROT. We excluded local dynamo action in the granular flow as the culprit for
the mismatch. The existence of a magnetic field for other reasons with a large
filling factor and sufficient strength (several 100\pun{G}) able to noticeably
influence the granular dynamics could qualitatively explain the observed low
level of granular brightness fluctuations. Whether this is viable from the
viewpoint of dynamo theory is unclear. In this context we discussed the
observational finding that the star, while magnetically rather active, does not
exhibit a magnetic field organized on scales comparable to the stellar radius
\citep{Catala+al09}.

Recently suggested new sets of fundamental stellar parameters
for HD\,49933 \citep{Kallinger+al08,Bruntt09} are unlikely to remove the
discrepancy -- primarily since they suggest a lower surface gravity than we
assumed in the present modelling. In contrast, when increasing the gravity by
0.2\pun{dex} (to about \logg=4.45), and decreasing the effective temperature
by 200\pun{K} (to about 6550\pun{K}) we can enforce a plausible, albeit not
perfect, correspondence between theory and observation.  Whether this is
compatible with constraints from stellar structure, 3D atmospheric modelling,
and p-mode excitation needs to be assessed. For the moment we are left with a
puzzling discrepancy between the predicted and observed granulation background
in HD\,49933, with only a rather ad-hoc idea for a remedy at hand.

The granulation-related brightness fluctuations are sensitive
to the stellar metallicity. A reduction of the metallicity from solar to 1/10
solar reduced the total fluctuation power by a factor of 0.58 at the studied
\Teff\ and \logg.   

If in the future one is able to devise an observational or theoretical
calibration of the granular background signal, its high sensitivity to the
surface gravity could render it an effective gravimeter.  \COROT\ has obtained
high-precision photometry for many giants. The stronger convection-related
background signal makes this easy to detect and quantify. We consider giants
as additional, interesting test cases for our understanding of the dynamics of
convective flows on the surface of stars.

\acknowledgements 
The authors thank Michel Auvergne for providing information on the change of
instrumental sensitivity, and Hans Bruntt for communicating his latest results
on the fundamental parameters of HD\,49933 before publication. HGL would like
to thank Claude Catala for discussions about spectroscopy and polarimetry of
HD\,49933 on several occasions. HGL acknowledges financial support from EU
contract MEXT-CT-2004-014265 (CIFIST)

\bibliographystyle{aa}
\bibliography{microvar2}

\end{document}